\documentstyle[prc,aps,epsfig]{revtex}

\begin{document}

\date{\today} 
\title{Measuring centrality with slow protons in
proton-nucleus collisions at the AGS}

\author{(The E910 Collaboration, Brookhaven National Laboratory)}
\author{
	I.~Chemakin$^{2}$,
	V.~Cianciolo$^{7,8}$,
	B.A.~Cole$^{2}$,
	R.~Fernow$^{1}$,
	A.~Frawley$^{3}$,
	M.~Gilkes$^{9}$,
	S.~Gushue$^{1}$,
	E.P.~Hartouni$^{7}$,
	H.~Hiejima$^{2}$,
	M.~Justice$^{5}$,
	J.H.~Kang$^{11}$,
	H.~Kirk$^{1}$,
	N.~Maeda$^{3}$,
	R.L.~McGrath$^{9}$,
	S.~Mioduszewski$^{10}$,
	D.~Morrison$^{10,1}$,
	M.~Moulson$^{2}$,
	M.N.~Namboodiri$^{7}$,
	G.~Rai$^{6}$,
	K.~Read$^{10}$,
	L.~Remsberg$^{1}$,
	M.~Rosati$^{1,4}$,
	Y.~Shin$^{11}$,
	R.A.~Soltz$^{7}$,
	S.~Sorensen$^{10}$,
	J.~Thomas$^{7,6}$,
	Y.~Torun$^{9,1}$,
	D.~Winter$^{2}$,
	X.~Yang$^{2}$,
	W.A.~Zajc$^{2}$,
	and Y.~Zhang$^{2}$,
	}

%
%	institutions
%
\bigskip
\address{
$^{1}$ Brookhaven National Laboratory, Upton, NY 11973\\
$^{2}$ Columbia University, New York, NY 10027 and Nevis Laboratories, Irvington, NY 10533\\
$^{3}$ Florida State University, Tallahassee, FL 32306, \\
$^{4}$ Iowa State University, Ames, IA 50010, $^{5}$ Kent State University, Kent, OH 44242\\
$^{6}$ Nuclear Science Division, Lawrence Berkeley National Laboratory, Berkeley, CA 94720 \\
$^{7}$ Lawrence Livermore National Laboratory, Livermore, CA 94550 \\
$^{8}$ Oak Ridge National Laboratory, Oak Ridge, TN 37831\\
$^{9}$ State University of New York at Stonybrook, Stonybrook, NY 11794\\
$^{10}$ University of Tennessee, Knoxville, TN 37996, $^{11}$ Yonsei University, Seoul 120-749, Korea \\
}

\maketitle

\pagebreak
\begin{abstract}

Experiment E910 has measured slow protons and deuterons from
collisions of 18~GeV/c protons with Be, Cu, and Au targets at the BNL
AGS.  These correspond to the ``grey tracks'' first observed in
emulsion experiments.  We report on their momentum and angular
distributions and investigate their use in measuring the centrality of
a collision, as defined by the mean number of projectile-nucleon
interactions.  The relation between the measured $N_{grey}$ and the
mean number of interactions, $\overline{\nu}(N_{grey})$, is studied
using several simple models, one newly proposed, as well as the RQMD
event generator.  RQMD is shown to reproduce the $N_{grey}$
distribution, and exhibits a dependence of $N_{grey}$ on centrality
that is similar to the behavior of the simple models.  We find a
strong linear dependence of $\overline{N_{grey}}$ on $\nu$, with a
constant of proportionality that varies with target.  For the Au
target, we report a relative systematic error for extracting
$\overline{\nu}(N_{grey})$ that lies between 10\% and 20\% over all
$N_{grey}$.

\end{abstract}

\pagebreak
\twocolumn

\section{Introduction}
\label{sec:intro}

The use of high-energy collisions of hadrons with nuclear targets to
study the space-time development of produced particles was first
suggested many years ago~\cite{Fei66,Dar72,Gol73,Fis74}.  Early
experiments indicated that at sufficiently high energies, the
projectile will undergo on average a number of inelastic hadron-nucleon
scatterings roughly equal to the mean interaction thickness,
$\overline{\nu}=A\sigma_{hp}/\sigma_{hA}$, with most particles forming well
outside the target nucleus~\cite{Bus75,Eli80,Yea77,Bus75:HEP75}.
These data suggest that a single p-A collision can effectively be
modeled by a cascade of $\nu$ proton-nucleon interactions, with
$\sim$1~fm formation times for produced particles.  For reasons
given below, any conflicts between such a cascade model and p-A
data have yet to be demonstrated.

The differences between a p-A collision and a p-nucleon cascade are
especially important to discover and understand in light of recent
experiments with relativistic heavy-ion collisions at BNL and CERN.
Here the complex hadronic physics processes that we wish to study in
p-A form a significant background in the search for a QCD phase
transition.  The overwhelming complexity of A-A collisions makes it
difficult to study these processes directly, whereas p-A collisions
are simpler and may provide more insight.

Many previous p-A experiments were limited by their inability to
trigger on central collisions, those with small impact parameter and
in which $\nu$ attains the highest values --- essential to studying
the effects of multiple interactions.  Other experiments which could
trigger on centrality were limited by low rates (low statistics)
and/or insufficient phase space coverage for identified particles.
However, they were able to establish a relationship between $\nu$ and
a measurable observable, the number of slow singly charged fragments
(grey tracks\cite{greynote}) emitted in the collisions.  This
relationship is expressed as a conditional probability for detecting
$N_{grey}$ grey tracks given a collision in which there were $\nu$
interactions, $P(N_{grey}|\nu)$.  Given a distribution $\pi(\nu)$ for
the number of interactions, the relevant quantity for measuring
centrality in p-A collisions is,
\begin{equation}
\overline{\nu}(N_{grey}) = \sum_{\nu} \nu P(N_{grey}|\nu) \pi(\nu).
\label{eq:meannu}
\end{equation}
Several forms have been proposed for $P(N_{grey}|\nu)$~
\cite{And78:PAnu,Ste82:PAnu,Heg81:PAnu,Heg82:PAnu,Suz82}, yet there
have been few systematic studies to test the validity of the models'
assumptions and assess the accuracy of the extracted values of
$\overline{\nu}(N_{grey})$.

We will focus on the two models which have been most commonly applied
to data: the Geometric Cascade Model (GCM) of Andersson~{\it et
al.}~\cite{And78:PAnu}, and the intra-nuclear cascade calculation of
Hegab and H\"{u}fner~\cite{Heg82:PAnu,Heg81:PAnu}.  We will then
present a new model which draws on elements of both.  The GCM uses a
normalized geometric distribution for $P(N_{grey}|\nu=1)$ and assumes
that this distribution applies equally and independently to the
distribution of grey tracks produced by each primary proton-nucleon
scattering.  This yields an analytic form for joint probability
distribution, $P(N_{grey},\nu)$, which has,
\begin{equation}
\overline{N_{grey}} \propto \overline{\nu}.
\label{eq:nulin}
\end{equation}
The calculation of Hegab and H\"{u}fner performs a sum over the
collisions of the beam and all primary struck nucleons, assuming a
straight line path through the nucleus for all products which follows
the initial impact parameter of the projectile.  The mean value of
this distribution is given approximately by,
\begin{equation}
\overline{N_{grey}} \propto \overline{\nu}^2.
\label{eq:nuquad}
\end{equation}
Despite this fundamental difference, both models and variations of
them have successfully reproduced the $P(N_{grey})$ distributions for
a number of experiments.
See~\cite{Bab78,And78:PAnu,Ree83,Dem84:PA,Bru85,Alb93}, for the GCM,
and~\cite{Heg82:PAnu,Bai87} for the cascade of Hegab and H\"{u}fner.
Values of $\overline{\nu}(N_{grey})$ have been extracted for many
types of experiments: emulsions~\cite{Bab78},
counters~\cite{Alb93,Bra82}, and streamer/bubble
chambers~\cite{Ree83,Dem84:PA,Bai87,Bri89:PA}.  The GCM model has also
been applied to the $N_{grey}$ distribution from $\nu$-Ne
interactions~\cite{Bru85}.  In each case, the agreement between model
and data is quite reasonable given the simplistic nature of the
models, but the accuracy of the $\overline{\nu}(N_{grey})$ extraction
is undetermined.  If the systematic errors are small compared to the
range of $\nu$ for a given target, then the analytic approaches to
determine $\overline{\nu}(N_{grey})$ are justified.

Here we present a high statistics analysis of low momenta protons from
collisions of 18~GeV/c protons incident on three nuclear targets: Be,
Cu, and Au.  The data were taken by BNL E910, a large acceptance TPC
spectrometer experiment with additional particle identification from
time-of-flight (TOF) and \v{C}erenkov (CKOV) detectors.  To extract
$\overline{\nu}(N_{grey})$ and assess its accuracy we apply several models
to these data and to the distributions produced by
RQMD~\cite{Sor89:RQMD}, a cascade model for p-A and A-A collisions.
We estimate the systematic errors inherent in the models and in
the assumptions of the definition of $N_{grey}$.

The E910 experiment is described in Sec.~\ref{sec:exp}.  In
Sec.~\ref{sec:data} we present the reduction of the data, including
all cuts and corrections.  Final results are shown in 
Sec.~\ref{sec:results}.  Sec.~\ref{sec:model} contains the comparisons
to RQMD, and we determine the systematic errors in
Sec.~\ref{sec:syserr}.  In Sec.~\ref{sec:conc} we present our
conclusions.  In all included figures, we will continue to use the
term ``$N_{grey}$'' to refer to the number of singly charged slow
fragments measured by our TPC in a collision, to be consistent with
most of the literature.  Other commonly used terms for the grey tracks are
``prompt protons'' and ``slow particles''.

\section{Experimental Overview}
\label{sec:exp}

\begin{figure}
  \begin{center}
    \includegraphics[width=6in]{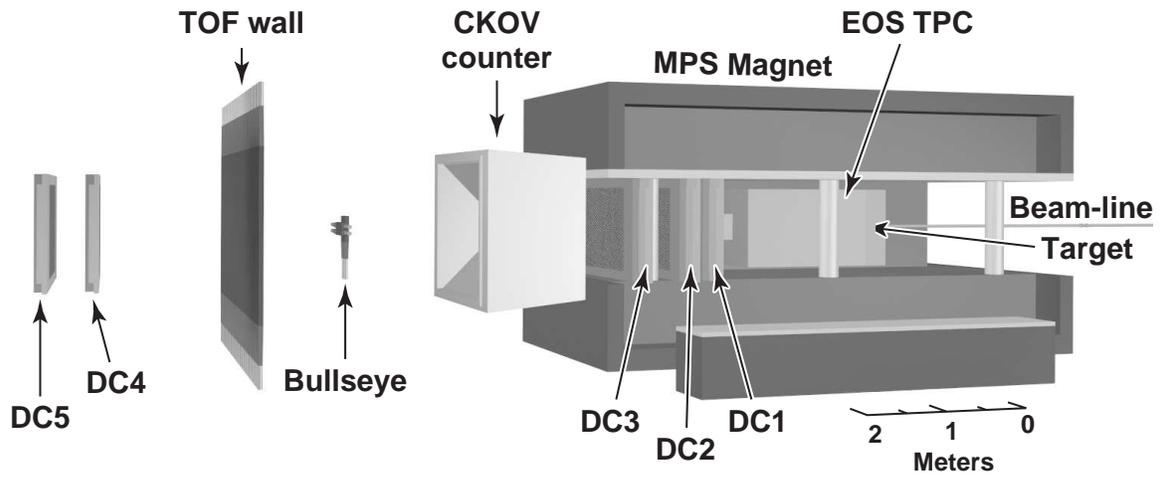}
    \caption{E910 layout.}
    \label{fig:e910}
  \end{center}
\end{figure}

The experimental layout for E910 is shown in Fig.~\ref{fig:e910}.  The
following discussion assumes a coordinate system that is right-handed
Cartesian, with the beam direction nominally along the z-axis and the
y-axis along the vertical.  The time projection chamber (EOS
TPC~\cite{Rai90:TPC}) has dimensions 96~x~75~x~154~cm, and is read out
through a 120~x~128 cathode-pad array.  The TPC was placed in the center
of the MPS magnet, which had a nominal central field of 0.5~T.  It ran
with P10 gas at atmospheric pressure with a vertical electric field of
120~V/cm.  Additional charged particle tracking immediately downstream
was provided by three drift chambers (DC1-3), placed near the end of the
magnet.  The drift chambers each had an active area of 172~x~100~cm,
with 7 planes each, consisting of three views in x (one staggered),
two in y (staggered), and two more views offset by $\pm 60^{\circ}$
from the vertical.  The \v{C}erenkov counter, with 139.7~x~190.5~cm
aperture, was filled with Freon~114 and placed 4.8~m downstream of
the target.  Two mirror planes, above and below the
vertical-mid-plane, with 48 mirrors each focused the light onto an
equal number of phototubes at the top and bottom of the counter.  The
TOF wall consists of 32 counters, each 15.2~x~178~x~4.8~cm arranged in
a flat panel, 610~x~370~x~86~cm, placed 8~m downstream and normal to
the z-axis.  Two more drift chambers (DC4-5) sat downstream of the TOF wall,
9.6 and 10.1~m from the target.  For these data, a bullseye
scintillator detector was placed between the \v{C}erenkov and TOF,
6.8~m from the target.  It consisted of two scintillators,
14.6~x~30.5~cm adjacent in x, and behind them two more of dimensions
40.6~x~7.6~cm adjacent in y.

Protons with nominal beam momenta of 6, 12, and 18~GeV/c were normally
incident on targets of Be, Cu, Au, and U.  Only the 18~GeV/c beam and
Be, Cu, and Au targets are included in this analysis.  The targets,
4\%~Be, 3\%~Cu and 2\%~Au targets were 3.9, 4.2, and 3.4~gm/cm$^2$
thick respectively, and were located in the TPC re-entrant window,
10~cm before the TPC active volume.  Beam
definition was provided by the S1 and ST scintillators.  S1 was placed
3.8~m upstream of the target.  It had dimensions 5~x~5~x~0.5~cm and
was read out by two phototubes on opposite sides.  ST, placed in front
of the target, provided the coincidence for the trigger.  It had
dimensions 10~x~10~x~0.1~cm and was readout by a single phototube.
Two veto counters, V1 and V2 were used to tune the beam and to reject
halo and upstream interactions.  V1 provided a 2~cm diameter circular
aperture 9~cm downstream of S1, and V2 provided a 2~x~1~cm rounded
aperture, 47~cm upstream of the target.  Beam vectoring was achieved
with two multi-wire chambers, A5 and A6, each with two horizontal and
two vertical views.  A5 was 10.36~m and A6 was 4.34~m upstream of the
target.  Four similar chambers with only X views, A1--A4, surrounded a
series of six dipole magnets further upstream to measure the average
beam momentum.  Just upstream of A5, three beam \v{C}erenkov
counters, C$_1$--C$_3$, were placed in the beamline to reject pions 
and kaons (C$_2$ only) in the beam.

E910 ran in the A1 secondary beam line of the AGS, with a typical
intensity of $3\cdot10^4{\rm s}^{-1}$.  For these data the beam momentum
was determined by A1-A4 reconstruction to be $17.5\pm0.2$(sys)~GeV/c.
The LVL0 trigger required a coincidence of ST and S1 (which
provided the start time for the experiment), in anti-coincidence with
the veto and beam \v{C}erenkov counters:
\begin{equation}
{\rm LVL0} = S_{1}\wedge ST\wedge \overline{V_{1}}\wedge \overline{V_{2}}\wedge
\overline{C_{1}}\wedge \overline{C_{2}}\wedge \overline{C_{3}}.
\label{eq:beamdef}
\end{equation}
The beam trigger definition furthermore required the absence of a
signal in S$_1$ during the preceding 1~$\mu s$.  Beam triggers with no
corresponding hit in the bullseye scintillator satisfied the
interaction trigger.  Final event statistics (after cuts) are given in
Table~\ref{tab:statsncuts}.  A sample of target-out
events were also taken.

\section{Data Reduction}
\label{sec:data} 

\begin{figure}
  \begin{center}
    \includegraphics[width=4in]{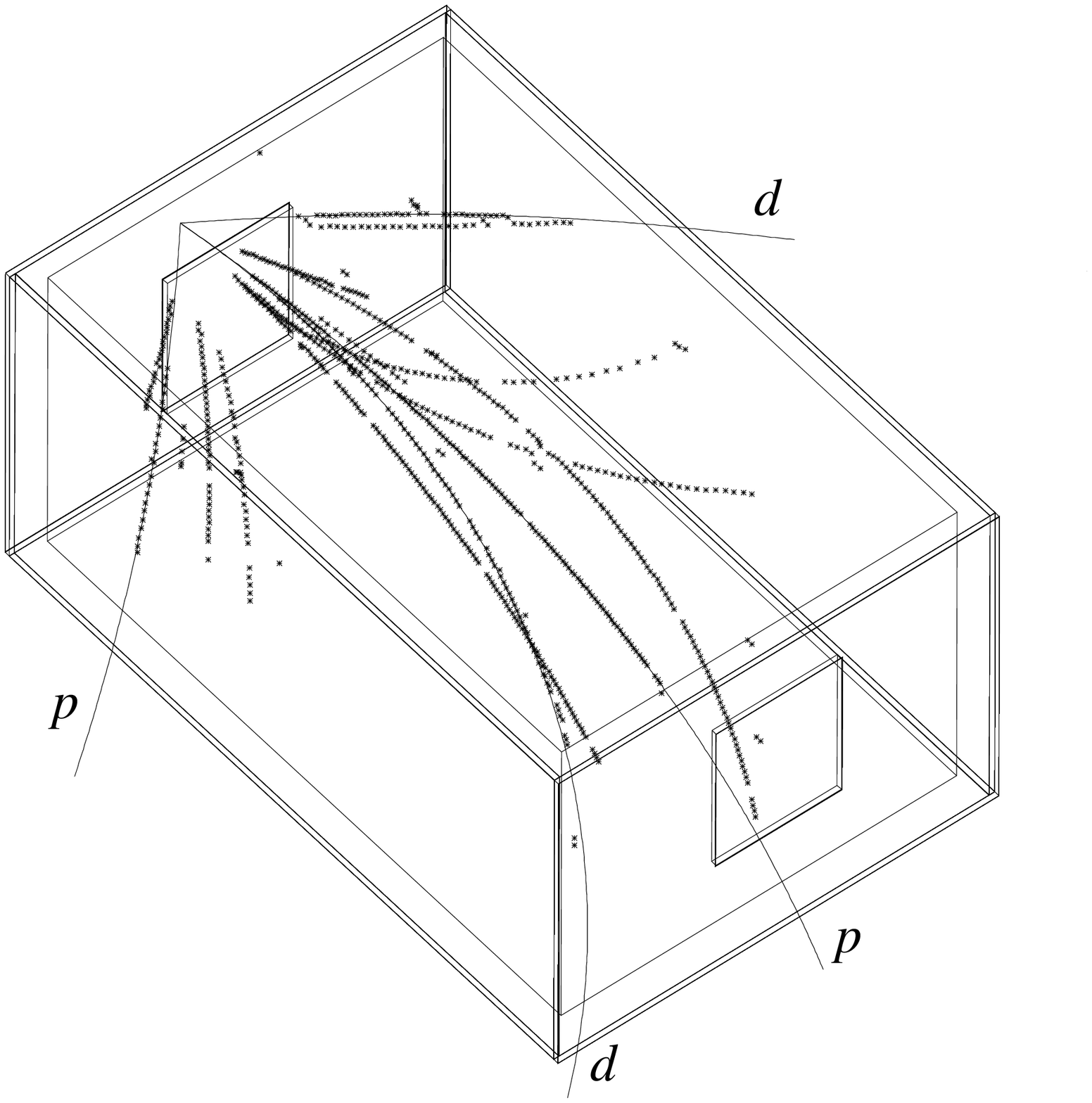}
    \caption{18~GeV/c p+Au event in EOS TPC.  Reconstructed tracks are
     drawn for protons and deuterons which contribute to $N_{grey}$.}
    \label{fig:tpcevent}
  \end{center}
\end{figure}

All particle tracking and identification to be presented here comes
from the TPC analysis described below.  Time and pulse-height
distributions are grouped into x-y clusters for each pad row using the
center for the z-coordinate.  A road-finding procedure extends the
clusters along either direction to form a track.  The initial momenta
are determined from a fit to a helix, assuming a constant dipole for
the field within the TPC and extending forward to the target.  Tracks
which originate from the target location ($\chi^2$ cuts are employed)
are used to determine the vertex.  Those tracks used in the vertex
determination are refit with fixed vertex to determine final momenta.
All tracks must pass appropriate $\chi^2$ cuts, have hits along ten or
more pad-rows in z, and originate from the event vertex to be included
in the $N_{grey}$ distribution.  A GEANT simulation of the TPC shows
the momentum resolution for the $N_{grey}$ tracks to be dominated by
multiple scattering, with a resolution of 15~MeV/c for 1~GeV/c
protons.  A typical event is shown in Fig.~\ref{fig:tpcevent}.

\begin{figure}
  \begin{center}
    \includegraphics[width=6in]{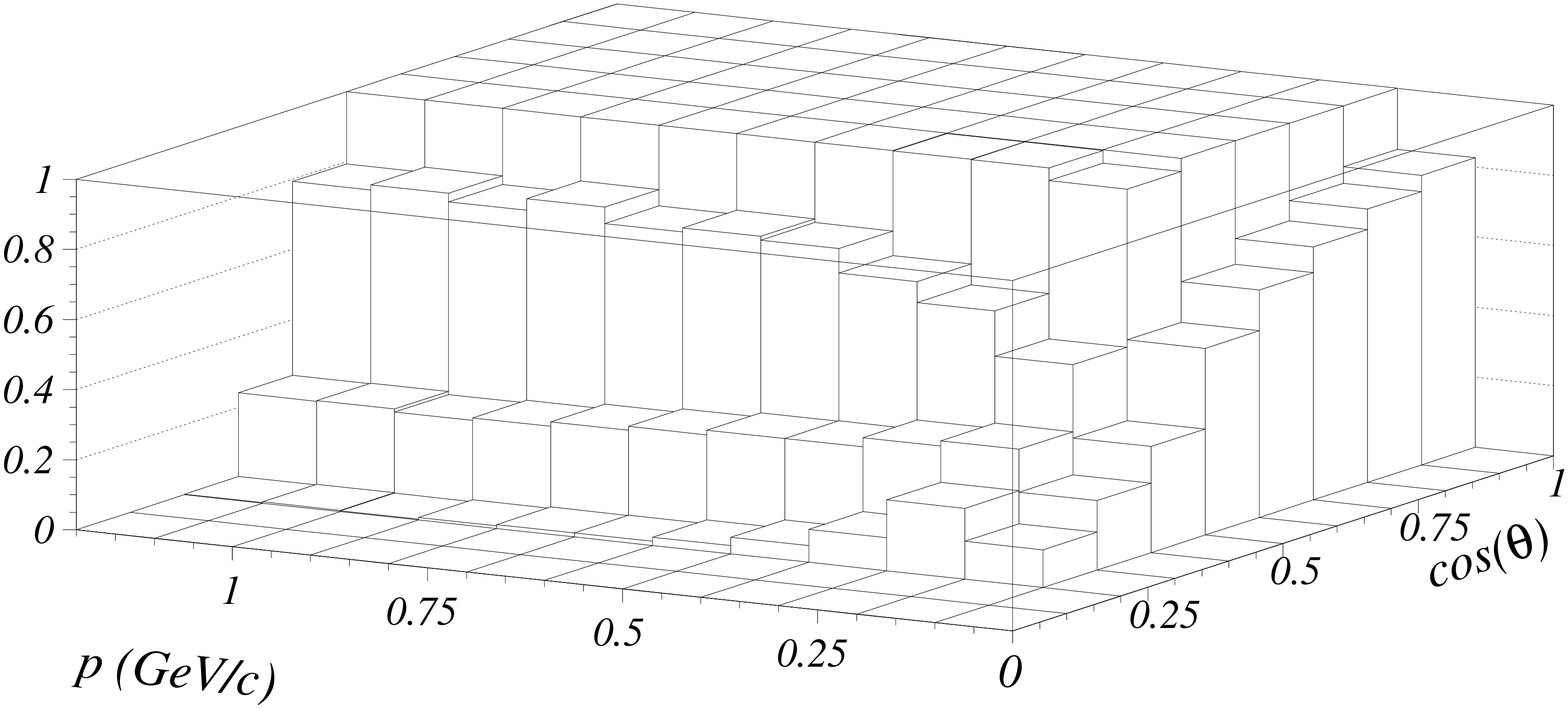}
    \vspace{0.5cm}
    \caption{
             TPC geometric acceptance as a function of momentum and
             $\cos(\theta)$.
	    }
    \label{fig:tpcaccept}
  \end{center}
\end{figure}

The TPC has good acceptance for the region forward of
$\cos(\theta)=0.4$ and above a momentum of 100~MeV/c.  The geometric
acceptance, shown in Fig.~\ref{fig:tpcaccept}, was calculated with
single track events thrown in a GEANT simulation with multiple coulomb
scattering enabled.  The full acceptance which accounts for
mis-reconstructed momentum extends the acceptance correction to the
lowest momentum bin, but in bins with finite geometric acceptance the
differences are less than 5\%. 

\begin{figure}
  \begin{center}
    \includegraphics[width=4in]{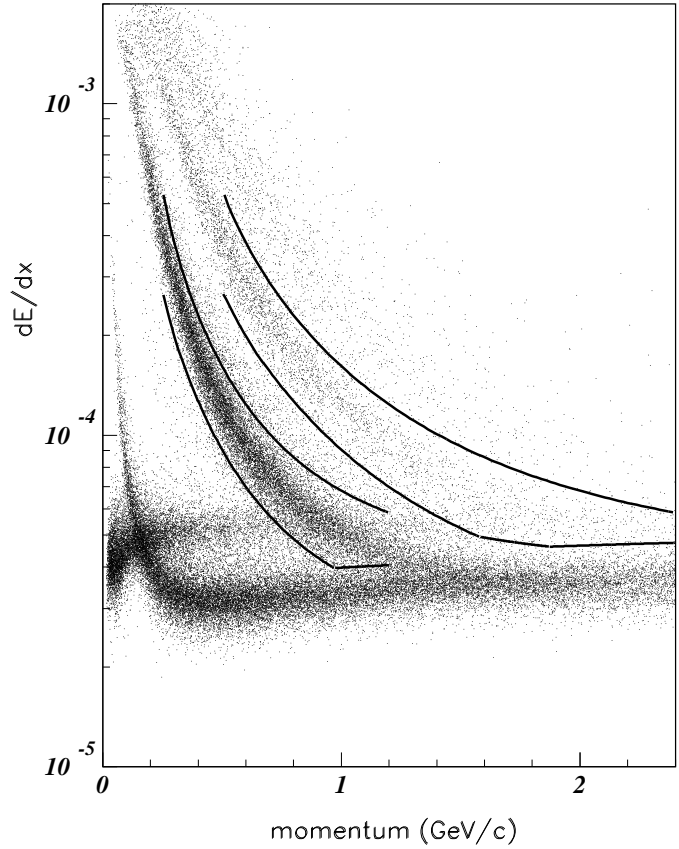}
    \caption{Ionization energy loss vs. momentum of particles with
      $p<2.4$~GeV/c.  The lines delimit the dE/dx particle
      identification cuts for protons and deuterons described in the text.}
    \label{fig:dedx}
  \end{center}
\end{figure}

Particles are identified in the TPC through their ionization energy
loss, dE/dx, calculated using a 70\% truncated mean.  The distribution
of dE/dx vs. momentum is shown is Fig.~\ref{fig:dedx}.  The dE/dx
distributions have been fit to the Bethe-Bloch formula with momentum
dependent gaussian widths.  This analysis does not correct for
saturation or non-linearities in the pulse-heights.  Particles with
dE/dx within $2.25\sigma$ of that for a proton and further than
$1.5\sigma$ from the pion dE/dx are identified as protons.  We require
that deuterons lie within $2.25\sigma$ of dE/dx for a deuteron and
further than $2.25\sigma$ from the proton and pion bands.  Protons are
identified up to a momentum of 1.2~GeV/c and deuterons up to
2.4~GeV/c.  Two additional cuts are required to limit positron
contamination coming from photon
conversions in the target (see Fig.~\ref{fig:dedx}).  Positive tracks
within the positron dE/dx 
band are matched to the negative track with a common vertex which
yields the smallest relative transverse momentum, $q_{T} = 2 \mid
\vec{p_{1}} \times \vec{p_{2}} \mid / \mid \vec{p_{1}} + \vec{p_{2}}
\mid$.  For $q_T < 0.037$~GeV/c, the positive track is removed from the
analysis.  From an application of this cut to a lower momentum region
we determined it to be $\sim$50\% effective for all targets.  Since
the positrons coming from $\pi^0$s (the dominant source of photons at
these momenta) are more forward peaked than low momentum protons and
deuterons, we furthermore reject positive tracks with dE/dx
consistent with that of a positron that are forward of
$\cos(\theta_f)$.  The value of $\cos(\theta_f)$, given in
Table~\ref{tab:statsncuts}, was chosen separately for protons and
deuterons for each target to minimize the contamination while
preserving statistics.  From the angular distributions of the
paired-positrons we estimate final positron contamination to be less
than 5\% of the overall $N_{grey}$ sample for all targets.

The bullseye interaction trigger accepts many elastic events and also
beam events in which the beam multiple coulomb scatters in the target
and TPC.  To remove these we require that an event contain two or more
charged particles emanating from the event vertex {\em or} a single
charged particle with transverse momentum greater than 0.06~GeV/c and
longitudinal momentum less than 12~GeV/c.  The reconstructed vertex
must lie within the projected x-y boundary of V2, and have a z-position
within 2.6~cm of the centroid for Au and Cu, and 1.75~cm for Be.  We
also require at least one hit in each view of A5 and A6
to reconstruct the beam vector for each event.  All momenta are
translated to the coordinate system aligned with the beam.  Final
$N_{grey}$ statistics are given in Table~\ref{tab:statsncuts}.

%
% Table tab:statsncuts goes here
%

A typical energy range used to select the $N_{grey}$ tracks is
$30<K.E.<400$~MeV~\cite{Alb93,Bab78} ($0.24<p<0.87$~GeV/c).  The
purpose of the lower bound is to reject fragmentation products.  That
of the upper bound is to reduce the contribution from primary struck
recoil protons.  We examine these cuts in light of recent
multi-fragmentation data.  The EOS collaboration has measured the
proton fragmentation spectra in nucleus-nucleus collisions for
1.2~GeV$\cdot$A~Au+C~\cite{Hau98:MF}.  The proton kinetic energy
spectra show clear evidence of a kink at 30~MeV, and were well fit
over the range 0--100~MeV by a two-component Maxwell-Boltzmann
distribution with slope parameters of $\sim$8~and~$\sim$50~MeV for the
lowest multiplicity events, which are most similar to p-A collisions.
The higher slope parameter is consistent with fits to spectra from
4~GeV/c~p+Pb in the range 40--150~MeV by a group at
KEK~\cite{Nak83:MF}.  For collisions that more closely resemble the
data presented here, fragmentation spectra for 1-19~GeV/c and
80-350~GeV/c p+Xe have been measured, but only for fragments with $Z
\geq 3$~\cite{Por89:MF,Hir84:MF}.  The fitted spectra in the range
10--100~MeV are consistent with the assertion that fragmentation
spectra should appear thermal, with a temperature set by the mean
Fermi momentum of the emitted fragments: $T =
\frac{1}{5}\frac{p^2_F}{M_N}$~\cite{Gol74:MF}.  Therefore, an
appropriate lower limit for $N_{grey}$ lies near the Fermi momentum,
in agreement with the typical lower momentum limits for $N_{grey}$
found in the literature.

\begin{figure}
  \begin{center}
    \includegraphics[width=6in]{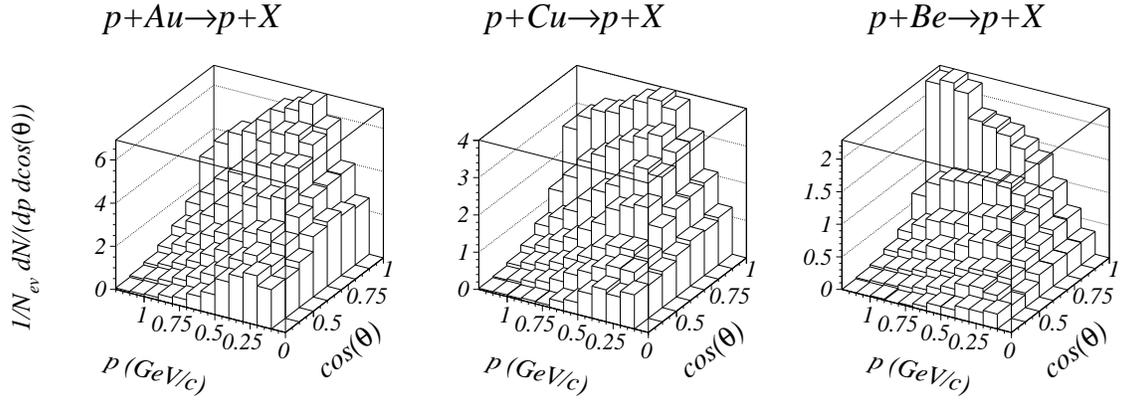}
    \caption{Acceptance corrected momentum {\it vs.} $\cos(\theta)$
     distribution for protons.}
    \label{fig:momcth}
  \end{center}
\end{figure}

\begin{figure} \begin{center}
  \includegraphics[width=6in]{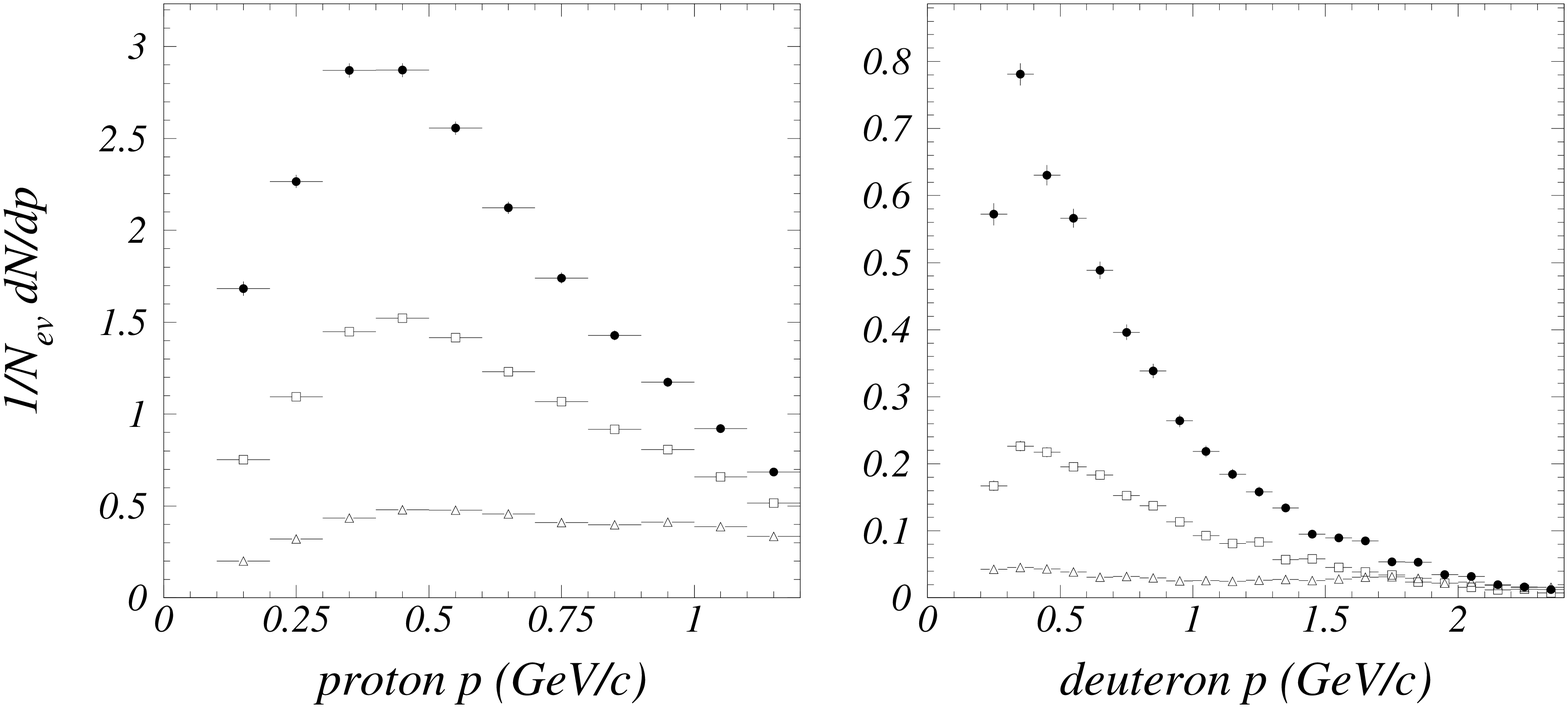}
  \caption{Acceptance corrected momentum distributions for protons and
  deuterons.  Black circles for Au, open squares for Cu, and open
  triangles for Be target.
  } \label{fig:mom} 
  \end{center}
\end{figure}

\begin{figure}
  \begin{center}
    \includegraphics[width=6in]{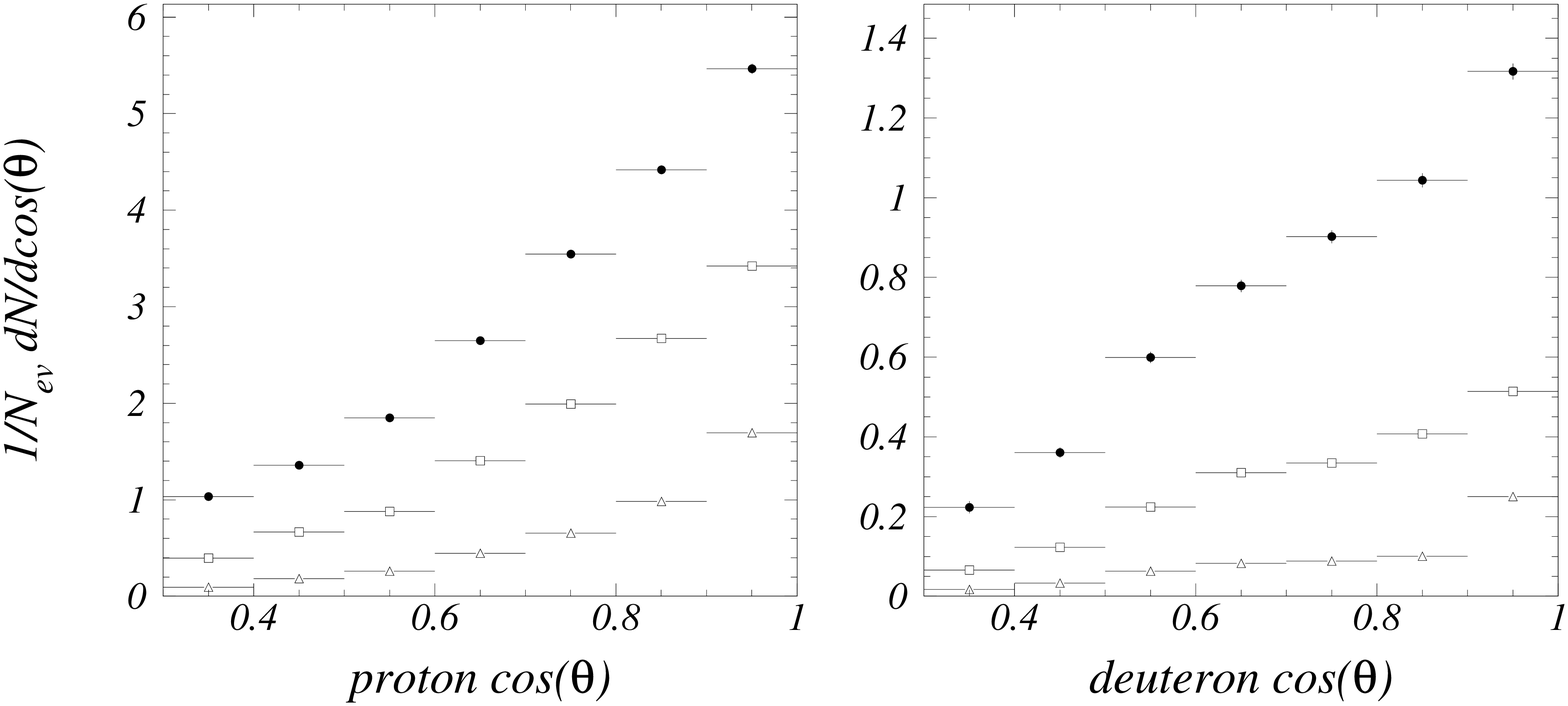}
    \caption{Acceptance corrected angular distributions for protons
    and deuterons.  Black circles for Au, open squares for Cu, and
    open triangles for Be target.} 
    \label{fig:cth}
  \end{center}
\end{figure}

Acceptance corrected momentum and angular distributions for protons
are given in Fig.~\ref{fig:momcth}.  Distributions are shown only for
$p>0.1$~GeV/c and $\cos(\theta)>0.3$, where the acceptance is greater
than 10\%.  The angular distributions for all targets are nearly
isotropic in the lowest bin, becoming progressively more forward
peaked at higher momenta.  The momentum distributions peak near
0.5~GeV/c for Au and at higher momentum for the lighter targets.
The projections in momentum and angle for both protons and
deuterons are shown in Fig.~\ref{fig:mom} and Fig.~\ref{fig:cth}.

\begin{figure}
  \begin{center}
    \includegraphics[width=6in]{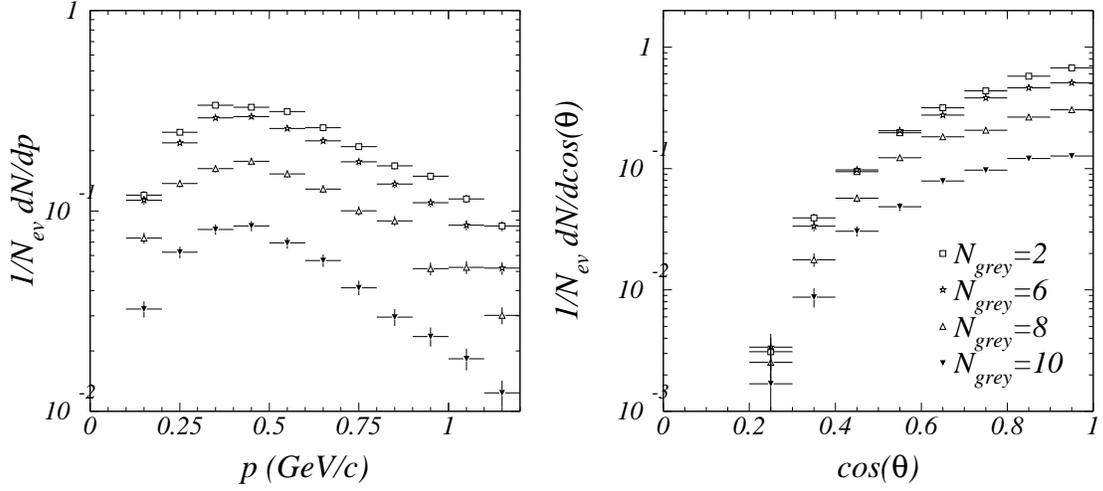}
    \caption{Momentum and angular distributions for values of 
    $N_{grey}=2,6,8,10$.}
    \label{fig:momcth_ng}
  \end{center}
\end{figure}

Based on these distributions, and the previous work in
multi-fragmentation, we use a range of $0.25<p<1.2$~GeV/c for protons,
and $0.5<p<2.4$~GeV/c for deuterons for our definition of $N_{grey}$.
The upper bounds reflect the limits of particle identification,
1.2~GeV/c for protons and 2.4~GeV/c for deuterons.  The upper limits
are higher than for most experiments, while the lower limits are
comparable.  We will explore the sensitivity of our analysis to this
choice of cuts in a study of the systematic errors presented in
Sec.~\ref{sec:syserr}.  With this definition of $N_{grey}$,
Fig.~\ref{fig:momcth_ng} shows the corrected momentum and angular
distributions for different values of $N_{grey}$ for the Au target.
The distributions do not shift backwards of the TPC acceptance for
large $N_{grey}$, an effect which would bias our determination of
$\nu$.

The distributions are corrected for target out contribution by
subtracting the beam normalized $N_{grey}$ distributions taken from
runs with an empty target holder.  After application of the vertex
cut, this correction amounts to 4\% (12\% of the $N_{grey}=0$ bin) 
for Au, and 2\% for Be and Cu.

\begin{figure}
  \begin{center}
    \leavevmode
    \includegraphics[width=6in]{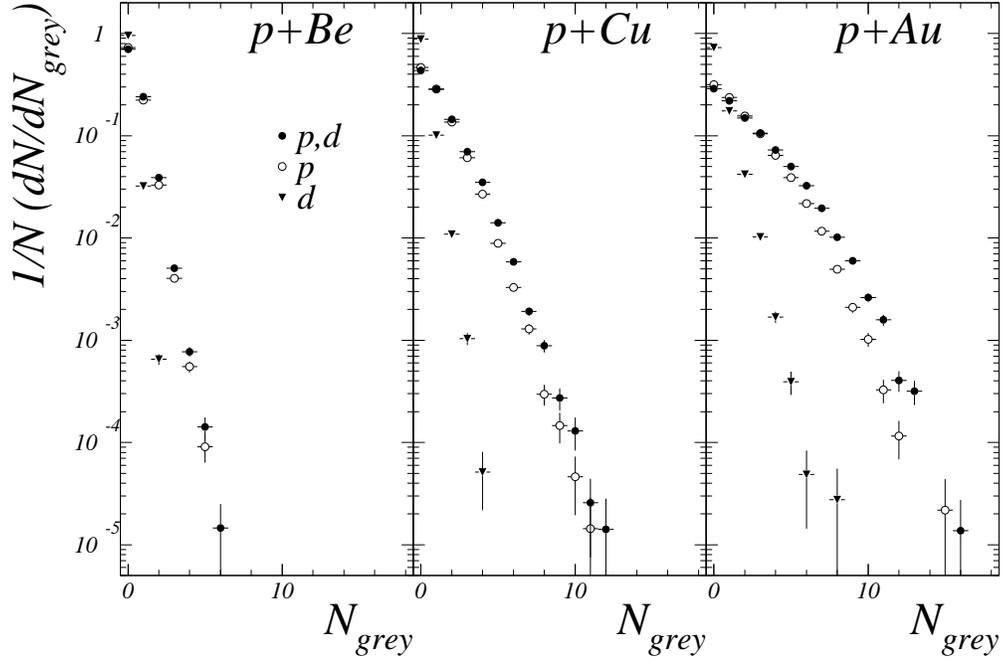}
    \caption{
      Event normalized multiplicity distributions
      $\frac{1}{N} dN/dN_{grey}$ of protons 
      (open circles), deuterons (triangles) and both protons and
      deuterons (dark circles).}
    \label{fig:nslow_final}
  \end{center}
\end{figure}

Finally, we correct for the contribution from secondary interactions
in the target (interactions of the projectile with a second nucleus).
The correction is performed iteratively, according to
Eq.~\ref{eq:reint},
\begin{eqnarray}
P_{n+1}(N_{grey}) & = & \frac{x_0}{x} e^{-(\frac{x}{x_0})}
P_{n}(N_{grey}) - \nonumber \\
& & \frac{1}{2}\frac{x}{x_0}
\sum_{i=0}^{N_{grey}}P_{n}(i)P_{n}(N_{grey}-i),
\label{eq:reint}
\end{eqnarray}
where $x_0$ is the p-A interaction length and $x$ is the interaction
thickness of the target.  Convergence is rapid and only a few
iterations are required.  Corrections for tertiary interactions have
been calculated and found to be negligible.  The final distributions
of slow protons and slow deuterons for all three targets are shown in
Fig.~\ref{fig:nslow_final}.

\section{Results}
\label{sec:results}

We begin with the GCM~\cite{And78:PAnu}, which assumes a normalized
geometric distribution of grey tracks for a single proton-nucleon
interaction:
\begin{equation}
P(N_{grey}|\nu=1) = (1-X) X^{N_{grey}}, \ \ \ X =
\frac{\mu}{1+\mu}
\label{eq:gcm_nu1}
\end{equation}
where $\mu$ is the average measured $N_{grey}$ when $\nu=1$.
Convoluting $\nu$ independent interactions,

\begin{equation}
P(N_{grey}|\nu) = \left(
\begin{array}{c} N_{grey} + \nu - 1 \\ \nu - 1 \end{array}  \right) 
(1-X)^{\nu}X^{N_{grey}}.
\label{eq:gcm_allnu}
\end{equation}
The resulting distribution is recognizable as a negative binomial,
where $\nu$ is the standard k-parameter, and the mean,
$\overline{N_{grey}}(\nu)$ is given by $\nu\mu$.  
Taking the weighted sum over $\nu$,
\begin{equation}
\overline{N_{grey}} = \sum_\nu \pi(\nu) \overline{N_{grey}}(\nu) =
\overline{\nu} \cdot \mu.
\label{eq:allgcmmean}
\end{equation}
Thus, Eq.~\ref{eq:nulin} is satisfied, a direct consequence of the sum
over $\nu$ independent distributions.  The full distribution is given
by, 
\begin{equation}
P(N_{grey}) = \sum_{\nu} P(N_{grey}|\nu) \pi(\nu).
\label{eq:gcm_ng}
\end{equation}

\begin{figure} 
  \begin{center} 
  \leavevmode
  \includegraphics[width=6.0in]{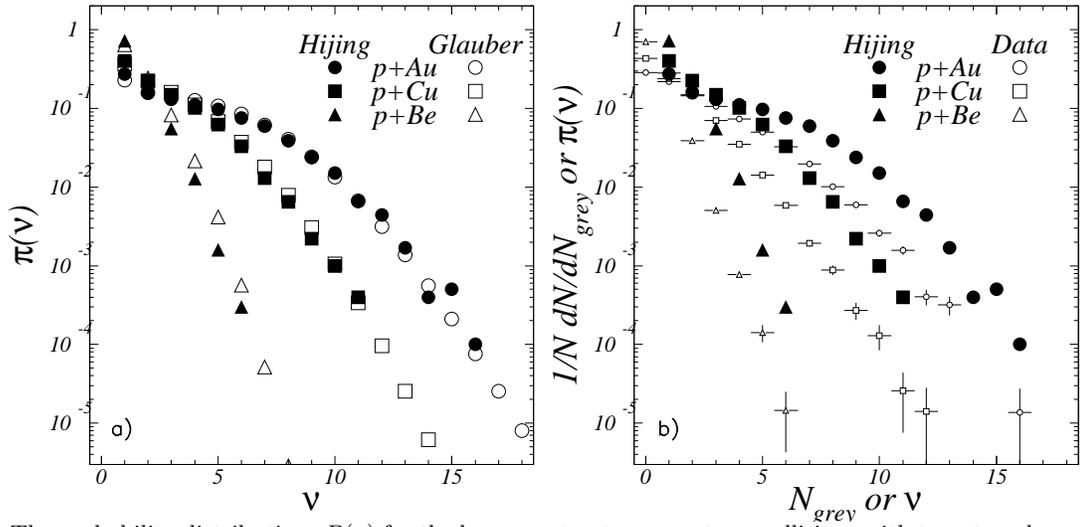}
  \caption{
  a) The probability distributions $P(\nu)$ for the
  beam proton to encounter $\nu$ collisions with target nucleons
  calculated for p+Be, p+Cu, and p+Au reactions using two different
  models: The Glauber results are based on the analytical Glauber
  model with Wood-Saxon nuclear density distributions and the Hijing
  results are based on a Monte-Carlo simulation of the collision
  geometry within the Hijing computer code. 
  b) The $\nu$
  distributions from Hijing overlayed with the $N_{grey}$
  distributions for all three targets.
  }
  \label{fig:nu_distributions} 
  \end{center}
\end{figure}

Two calculations for $\pi(\nu)$, Glauber and Hijing, are shown in
Fig.~\ref{fig:nu_distributions}a for the three targets.  Both
calculate $\nu$ from an optical model~\cite{Gla67} using a value of
30~mb for the p-N cross-section and a Wood-Saxon distribution of the
nucleus.  The Glauber calculation performs a numerical integration
over impact parameter (b) assuming a binomial probability distribution
for $\nu(b)$, where the mean and maximum values are given by the
nuclear thickness.  The results labeled Hijing~\cite{Wan91:Hijing}
come from the HIJING Monte-Carlo event generator which in this context
is equivalent to the LUND geometry code.  The two distributions are
similar.  We use the Hijing distribution for all further analysis
unless explicitly stated otherwise.  Fig.~\ref{fig:nu_distributions}b
overlays the $\pi(\nu)$ distributions with the measured $N_{grey}$
distributions.  The similarity between them is what prompted the
authors of~\cite{Bab78} to suggest that the $\nu$ and $N_{grey}$
distributions are correlated.

\begin{figure}
  \begin{center}
    \leavevmode
    \includegraphics[width=6.0in]{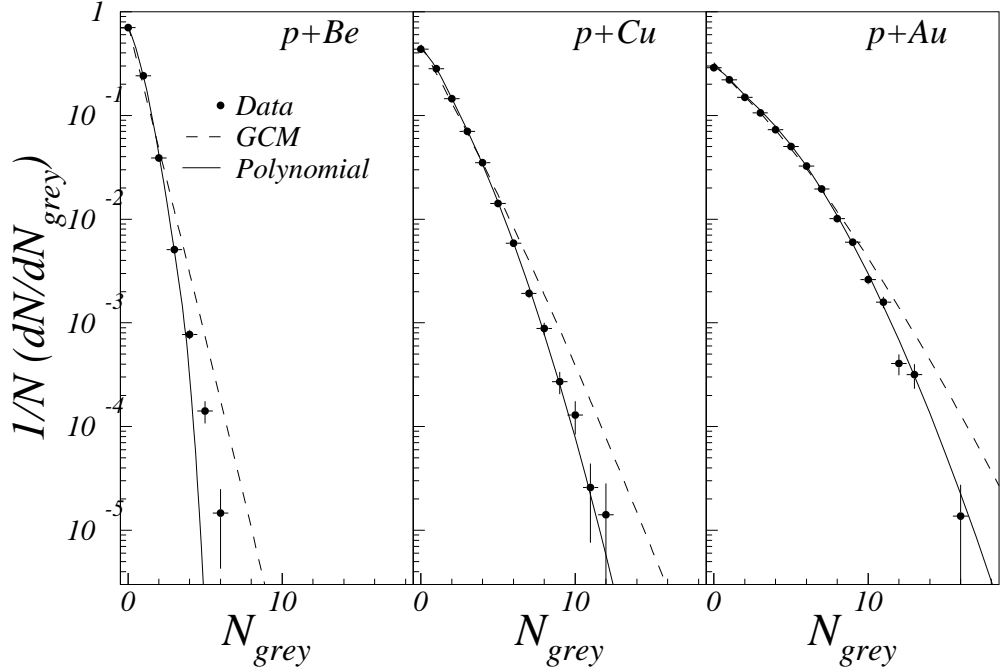}
    \caption{
      Log-likelihood fits to the event normalized $N_{grey}$
      distributions for Be, Cu, and Au targets with 
      two models: The Geometric Cascade Model (dashed lines) and the Polynomial
      Model (solid lines). 
      } 
    \label{fig:model_fits}
  \end{center}
\end{figure}

\begin{figure}
  \begin{center}
    \leavevmode
    \includegraphics[width=6.0in]{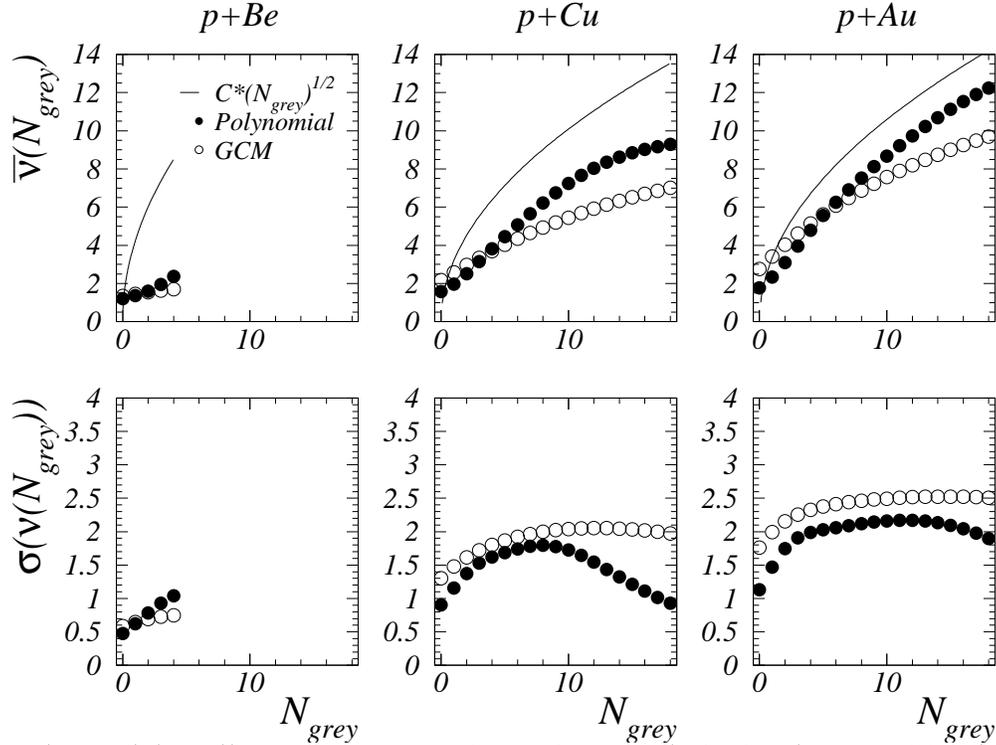}
    \caption{
      $\overline{\nu}(N_{grey})$ and $\sigma(\nu(N_{grey}))$  
      generated from the Polynomial Model (solid circles) and the GCM
      (open circles), and 
      $\overline{\nu}(N_{grey})$ according to the 
      $\overline{\nu}^2$ ansatz (solid line) for p+Be, p+Cu and p+Au.
      } 
    \label{fig:nu_Ng_datacomp}
  \end{center}
\end{figure}

The parameter $X$ in Eq.~\ref{eq:gcm_nu1} is related to the mean value
of $N_{grey}$ for a single proton-nucleon interaction, prompting many
authors to attempt to isolate the class of $\nu=1$ events through
multiplicity and leading particle cuts to determine $X$.  In the
context of the GCM, $\overline{N_{grey}}(\nu=1)$ is equal to the ratio
$\overline{N}_{grey}/\overline{\nu}$.  We follow the
method of~\cite{Alb93} and allow $X$ to be a free parameter in the fit
of the $N_{grey}$ distributions.  The results of the fits are given in
Table~\ref{tab:andfit}, along with the mean values, and ratio of
$N_{grey}$ and $\nu$.  Although for Au, the fitted X differs by
2$\sigma$, the fitted values of X for the other targets are identical
to the definition of X in Eq.~\ref{eq:gcm_nu1}.  The GCM fits are
displayed as the dashed curves in Fig.~\ref{fig:model_fits}.  The
model tends to fall below the data for low $N_{grey}$, and above the
data for high values.  This is reflected in the large values of
$\chi^2$/dof.  Note that the GCM distribution imposes no maximum on
the number of protons that can be emitted from a single nucleus.  The
mean and dispersion for $\nu$ are given by the probability
distribution in Eq.~\ref{eq:gcm_allnu}, displayed as the open circles
in Fig.~\ref{fig:nu_Ng_datacomp}.

%
% Table tab:andfit goes here
%

The intra-nuclear cascade of ~\cite{Heg81:PAnu,Heg82:PAnu,Cha83} takes
a very different approach in relating $N_{grey}$ to $\nu$.  It assumes:
\begin{enumerate}
 \item all primary struck nucleons follow the initial projectile
 trajectory,
 \item only secondary nucleons and a fraction (approximately one half)
       of the primary protons contribute to $N_{grey}$.
\end{enumerate}
The full cascade calculation is solved numerically~\cite{Heg82:PAnu},
but it has the feature that $\overline{\nu}$ is very nearly proportional to
$\sqrt{\overline{N_{grey}}}$.  From this, the authors make the following
ansatz~\cite{Heg81:PAnu},
\begin{equation}
\overline{\nu}(N_{grey}) = \overline{\nu} \sqrt{N_{grey} /
                                          \overline{N_{grey}}}. 
\label{eq:hegab}
\end{equation}
Applying Eq.~\ref{eq:hegab} leads to the solid curves in
Fig.~\ref{fig:nu_Ng_datacomp}, which differ significantly from the
predictions of the GCM.  Furthermore, the quadratic dependence of
$N_{grey}$ on $\nu$ is very different from the linear relationship
of the GCM.

The contradictory nature of these two models led us to introduce
another model, which allows
for both a linear and quadratic dependence of $N_{grey}$ on $\nu$,
with the relative strengths determined by a fit to the data.  The
principal assumption is that for a given target, there exists a
relation between the mean number of grey tracks detected and the
number of primary interactions which takes the form of a second degree
polynomial,
\begin{equation}
\overline{N_{grey}}(\nu) = c_{0} + c_{1}\nu + c_{2}\nu^{2}.
\label{eq:polynu}
\end{equation}
We furthermore assume that the
distribution is governed by binomial statistics; a total of Z target
protons exist which can be emitted and detected with probability
$\overline{N_{grey}}({\nu})/Z$, 
\begin{eqnarray}
P(N_{grey}|\nu) & = & \left(
\begin{array}{c} Z \\ N_{grey} \end{array}  \right) 
\left( \overline{N_{grey}}(\nu) \over Z \right) ^{N_{grey}} \times 
\nonumber \\
 & & \left(1 - {\overline{N_{grey}}({\nu}) \over Z}\right) ^{Z-N_{grey}}.
\label{eq:polyallnu}
\end{eqnarray}
The full distribution of $P(N_{grey})$ is again given by a weighted
sum over $\pi(\nu)$ of Eq.~\ref{eq:gcm_ng}, and the coefficients of
Eq.~\ref{eq:polynu} are derived from a fit to the data.  The fitted
function for this {\it polynomial} model is shown as the
solid curve in Fig.~\ref{fig:model_fits}, and the coefficients are
given in Table~\ref{tab:polyfit}.
The quadratic coefficients for both the Au and Cu targets were
determined to be zero.  For the Be target, the distribution does not
extend far enough to allow independent determination of a linear and
quadratic coefficient.  Given that in the fits to heavier targets the
linear term is dominant and the quadratic term is negligible, we
remove the quadratic component for the fits to the Be data.

%
% Table tab:polyfit goes here
%

Fig.~\ref{fig:model_fits} shows that the polynomial model reproduces
the data more accurately than the GCM.  For a negligible quadratic
term, the polynomial model differs from the GCM in only two respects,
the presence of a constant term in Eq.~\ref{eq:polynu} and the use of
binomial statistics.  The latter is a natural choice, which conserves
I$_z$ for the nucleons, but we have not given a physical motivation
for the constant term.  To check that its inclusion does not alter the
overall preference of the polynomial fit for a linear dependence of
$\overline{N_{grey}}$ on $\overline{\nu}$, we removed the constant
term and refit the data.  The parameters are listed in
Table~\ref{tab:polyfit2}.  The resulting quadratic terms are still
negligible, though finite, and the linear term remains the dominant
contribution for $N_{grey}$, even for large values of $\nu$.

%
% Table tab:polyfit2 goes here
%

\begin{figure}
  \begin{center}
    \leavevmode
    \includegraphics[width=6.0in]{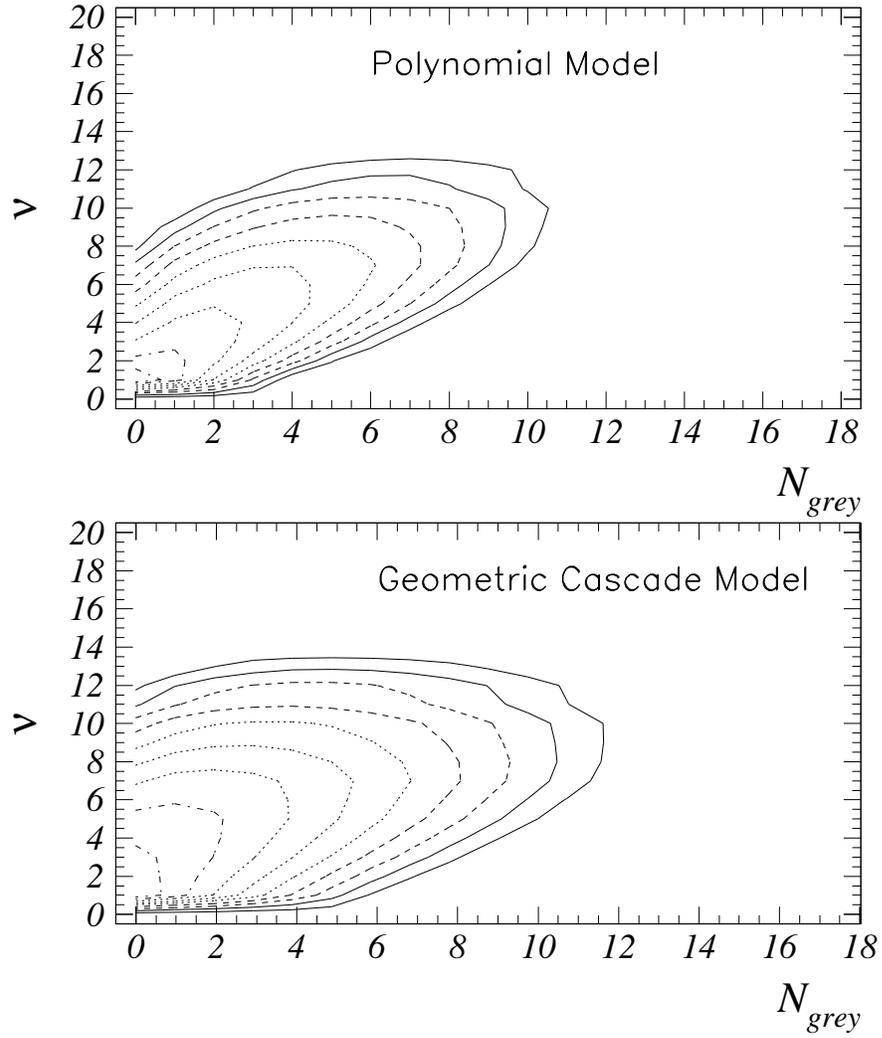}
    \caption{
      $P(N_{grey},\nu)$ contours for two models:  
      The Polynomial Model (top) and the Geometric Cascade Model
      (bottom). The ten contours for each are separated by factors of
      $\sqrt{10}$, ranging from 0.0001 to 0.316.
      } 
    \label{fig:nslow_vs_nu_2d}
  \end{center}
\end{figure}

Fig.~\ref{fig:nu_Ng_datacomp} gives $\overline{\nu}(N_{grey})$ for
all three models, and the dispersions for the GCM and polynomial
models.  The polynomial and GCM results are quite similar; they seldom
differ by more than 15\%, and never more than the dispersion of the
GCM.  In contrast, the intra-nuclear cascade differs significantly
from the other two, with the difference increasing for the lighter
targets.  The joint distributions, $P(N_{grey},\nu)$, for p-Au are
shown Fig.~\ref{fig:nslow_vs_nu_2d}.  Here the increased dispersion
for the GCM is evident, but otherwise the distributions again appear
quite similar.

\section{Model Comparisons}
\label{sec:model} 

Intra-nuclear cascade models have improved significantly since the
work of Andersson~{\it et al.} and Hegab and H\"{u}fner, and are now
capable of following the entire collision history in the context of
the classical approximations on which they are based.  There are now
several such models in the relevant energy range which have reproduced
many features of the available data for hadron-hadron, hadron-nucleus,
and nucleus-nucleus collisions.  These models will ultimately provide
a more accurate way to extract $\overline{\nu}(N_{grey})$, however
the large number of input parameters and assumptions require careful
study.  The aim of this section is to use one such model, RQMD, to
study the implications of the GCM and polynomial models.  This
provides an additional test of the systematic errors for these
models.  Their application to a newer cascade model also gives
an important historical point of reference.

RQMD (Relativistic Quantum-Molecular Dynamics) is a semi-classical
cascade model for hadron-nucleus and nucleus-nucleus
collisions~\cite{Sor89:RQMD}.  At AGS energies it functions primarily
as a transport code for the nucleons, excited nucleons, and produced
hadrons.  Particles can also interact through a mean field, here
disabled, and string formation, rare at these energies.  RQMD does not
simulate the nuclear fragmentation, and deuterons require the
additional application of a coalescence calculation.  The $N_{grey}$
count from our RQMD simulations includes only protons.  Presumably
some protons that contribute to $N_{grey}$ would bind with neutrons to
form deuterons with roughly twice the momentum of the proton.  These
deuterons would then fall within the $N_{grey}$ momentum range for
deuterons, leaving the overall $N_{grey}$ unaltered.

\begin{figure}
  \begin{center}
    \leavevmode
    \includegraphics[width=6.0in]{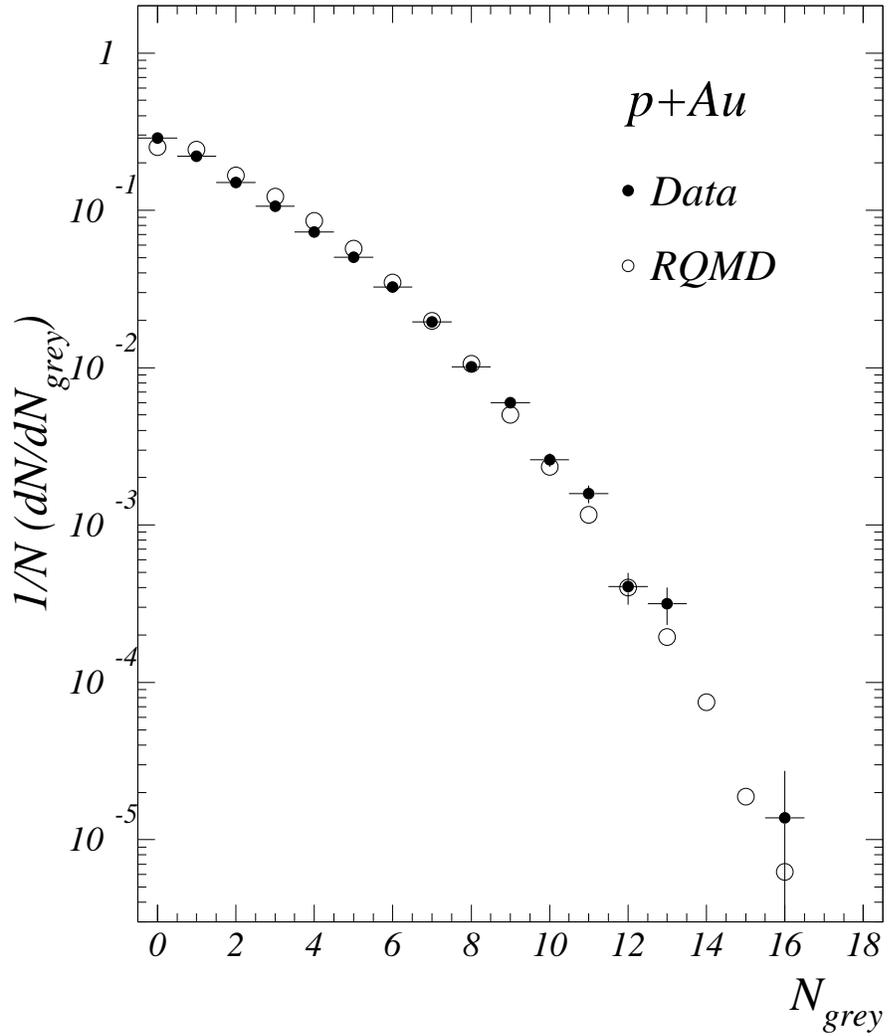}
    \caption{
      Comparison between event normalized slow fragment multiplicity
      distributions for p+Au reactions obtained from the E910 data and RQMD
      Monte-Carlo calculations.
      } 
    \label{fig:rqmd_nslow}
  \end{center}
\end{figure}

\begin{figure}
  \begin{center}
    \leavevmode
    \includegraphics[width=6.0in]{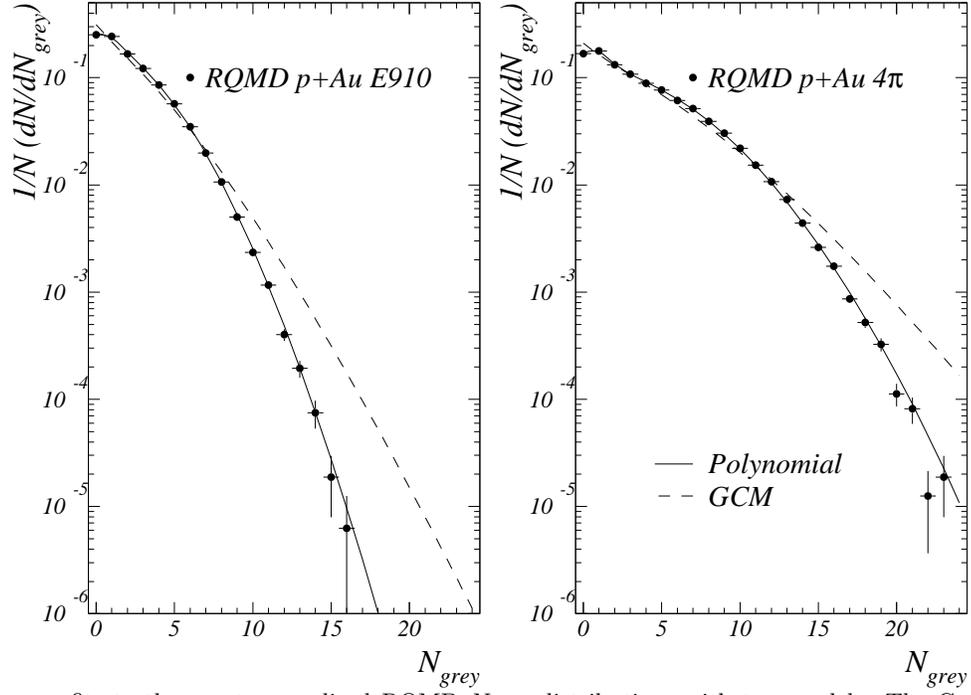}
    \caption{
      Chi-Square fits to the event normalized RQMD $N_{grey}$
      distributions with two models: The Geometric Cascade Model
      (dashed) and the Polynomial Model (solid), for both E910 and
      4$\pi$ model sets. 
      }
    \label{fig:rqmd_ngrey_fits}
  \end{center}
\end{figure}

A model data set of 200~K p+Au interaction events were generated with
RQMD 2.2 running in fast cascade mode in the fireball approximation
with all strong decays enforced.  The RQMD output was
then passed as input to the same GEANT simulation and track
reconstruction used to calculate the E910 acceptance.  The same
momentum cuts were used to define the grey tracks, although proton
identification was taken directly from the input.  We did not simulate
the positron contamination and no forward angle cuts were applied.
This data set is labeled ``RQMD E910''.  We also examine the full
distribution of $N_{grey}$ (no acceptance cuts), which includes all
protons within the momentum range specified for $N_{grey}$.
We refer to this data set as ``RQMD 4$\pi$''.  The $N_{grey}$ 
distribution for RQMD E910 is shown in Fig.~\ref{fig:rqmd_nslow},
along with the 
$N_{grey}$ distribution for the data (protons plus deuterons).  We see
that RQMD over-predicts the middle region of $N_{grey}$ and under-predicts
the extremes but nevertheless provides a reasonable description of the
data.
The GCM and polynomial fits were performed for both the RQMD E910 and
RQMD 4$\pi$ $N_{grey}$ distributions.  The analysis procedure remains
the same as it was for the data; the Hijing distribution for
$\pi(\nu)$ is again used in the fit.  The fitted functions are shown
in Fig.~\ref{fig:rqmd_ngrey_fits}, and the parameters are listed in
Table~\ref{tab:rqmd_ngrey_fits}.  The $X$ value obtained for RQMD
is larger than for the E910 data fits.  For the $4\pi$ set $X$
is larger by 35\% from additional protons which fall outside the E910
acceptance.  As with the data, the polynomial model gives a better
description of the $N_{grey}$ distributions for both E910 and $4\pi$
sets.  The quadratic coefficients are small and negative, but are not
consistent with zero, as was seen for the data.

%
% Table tab:rqmd_ngrey_fits goes here
%

\begin{figure}
  \begin{center}
    \leavevmode
    \includegraphics[width=6.0in]{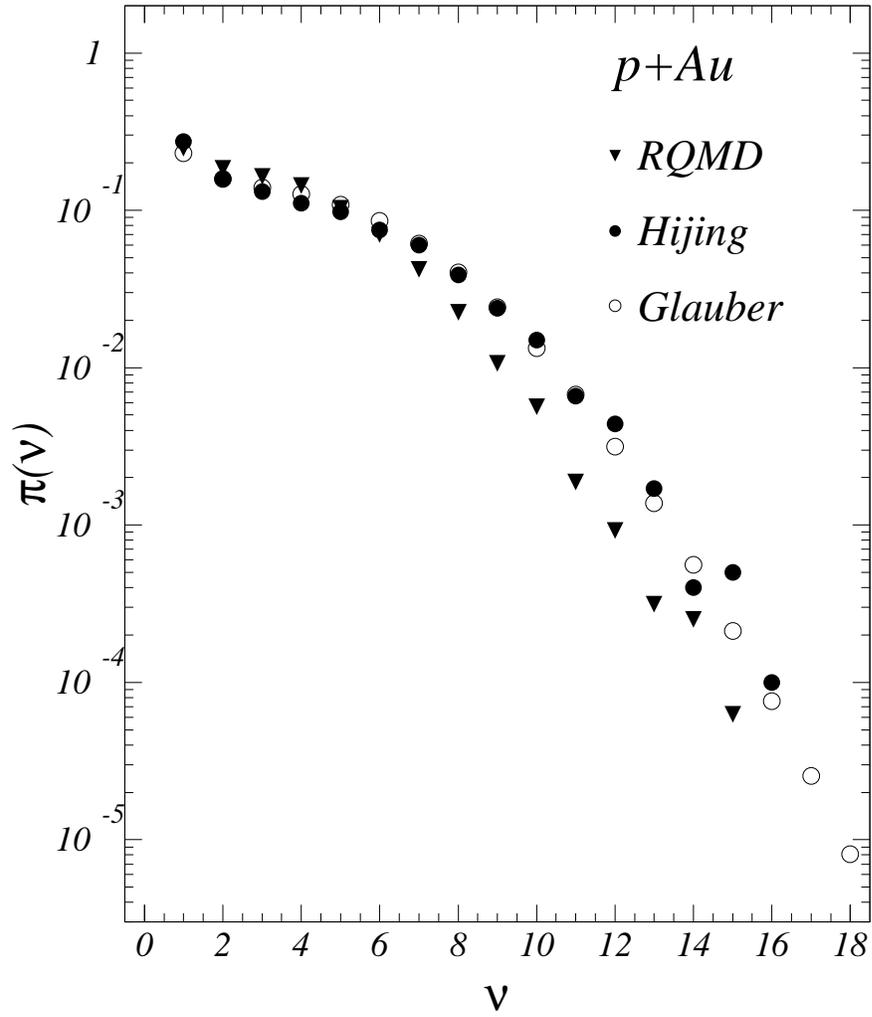}
    \caption{
      Comparison between the $\pi(\nu)$ distributions for p+Au reactions
      obtained with 3 different models: Glauber, Hijing and RQMD. 
      } 
    \label{fig:rqmd_nu}
  \end{center}
\end{figure}

The main goal in analyzing RQMD with the GCM and polynomial model is
to compare the extracted $\overline{\nu}(N_{grey})$ values with the
intrinsic $\nu$ of RQMD.  For RQMD, the definition of $\nu$ requires
some explanation.  Above a certain energy threshold, cross-sections in
RQMD are governed by the Additive Quark Model (AQM).  A hadron which
has one of its valence quarks assigned to a produced pion will have
its cross-section immediately reduced by 1/3, to be restored after a
proper time of 1~fm/c has passed.  Therefore the distribution of the
number of collisions reported for the projectile in the RQMD particle
file falls well below distributions for Glauber and Hijing shown
earlier.  To obtain the appropriate value of $\nu$ for comparison, we
examine the history file and count all collisions suffered by
particles that carry valence quarks of the projectile.  Counting for
produced particles stops when the formation times elapse, and multiple
collisions of valence quark-bearing particles with the same target
nucleon are counted only once.  The distribution of $\nu$ calculated
in this way is shown in Fig.~\ref{fig:rqmd_nu}, along with the Glauber
and Hijing calculations.  We see that for large $\nu$, RQMD falls
substantially below Glauber and Hijing.  It is interesting to note that the
relation of RQMD to Hijing in $\pi(\nu)$ is similar to its relation to
the data in the $N_{grey}$ distribution (see Fig.~\ref{fig:rqmd_nslow}).

\begin{figure}
  \begin{center}
    \leavevmode
    \includegraphics[width=6.0in]{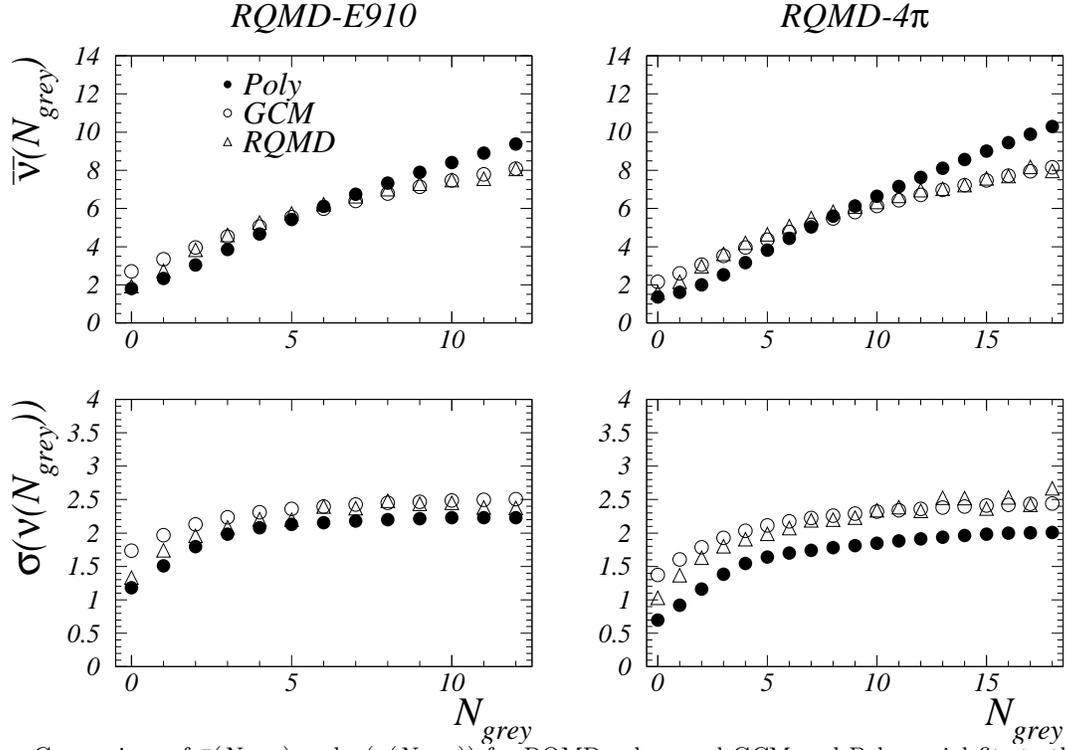}
    \caption{
      Comparison of $\bar{\nu}(N_{grey})$ and $\sigma(\nu(N_{grey}))$
      for RQMD values and GCM and Polynomial fits to the RQMD
     $N_{grey}$ within the E910 acceptance and over 4$\pi$.}
    \label{fig:rqmd_nu_ngrey}
  \end{center}
\end{figure}

\begin{figure}
   \begin{center} 
   \leavevmode
   \includegraphics[width=6.0in]{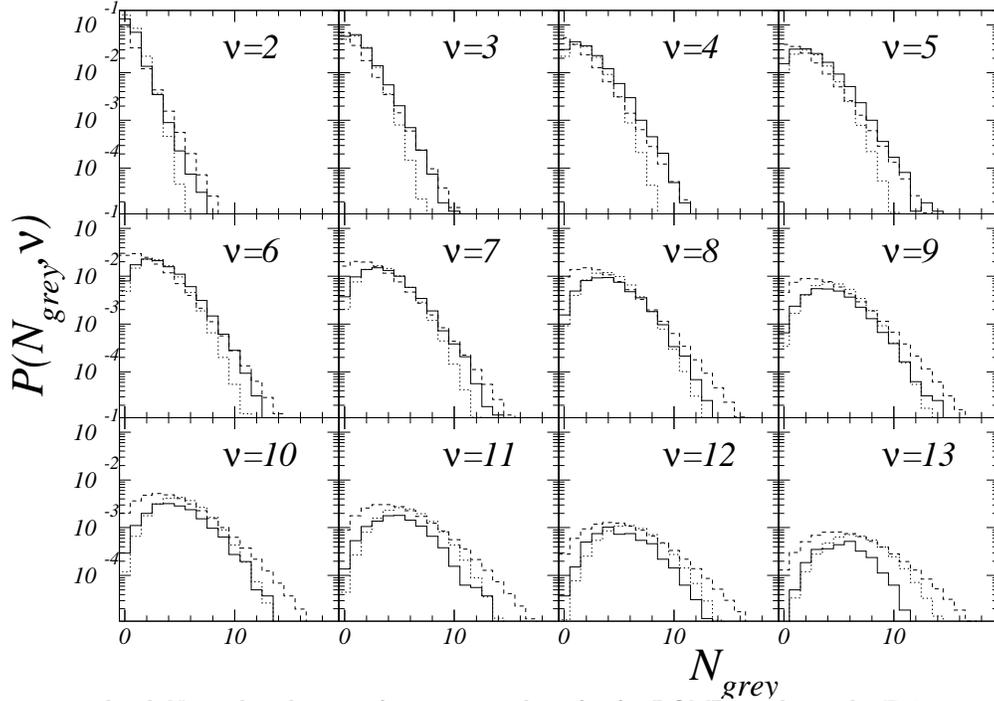}
   \caption{ 
    Event normalized $N_{grey}$ distributions for a given value
    of $\nu$ for RQMD tracks in the E910 acceptance, overlayed with the
    predictions of the two models --- dashed for the GCM, and dotted for
    the Polynomial.}  
  \label{fig:rqmd_ngrey_nuslices} 
  \end{center}
\end{figure}

The comparison for $\overline{\nu}(N_{grey})$ and
$\sigma(\nu(N_{grey}))$ among the GCM and polynomial analyses of RQMD
and their intrinsic values in RQMD are shown in
Fig.~\ref{fig:rqmd_nu_ngrey}. The GCM and polynomial models generally
differ by no more than one from the RQMD values in their prediction of
$\overline{\nu}(N_{grey})$.  The intrinsic RQMD values are matched by
the polynomial for the lowest $N_{grey}$, and by the GCM for
$N_{grey}>$3.  The intrinsic dispersions are bounded by the
predictions of the polynomial model below and the GCM above.  
It is also instructive to
examine $P(N_{grey},\nu)$ in slices of $\nu$, shown in
Fig.~\ref{fig:rqmd_ngrey_nuslices}.  The overall normalizations follow
the behavior of Fig.~\ref{fig:rqmd_nu}.  RQMD is
above the GCM and polynomial distributions for small $\nu$, and below
them for high $\nu$.  The $N_{grey}$ distributions for a given $\nu$
for RQMD are more accurately described by the polynomial model.

\section{Systematic Errors}
\label{sec:syserr} 

\begin{figure}
  \begin{center}
    \leavevmode
    \includegraphics[width=6.0in]{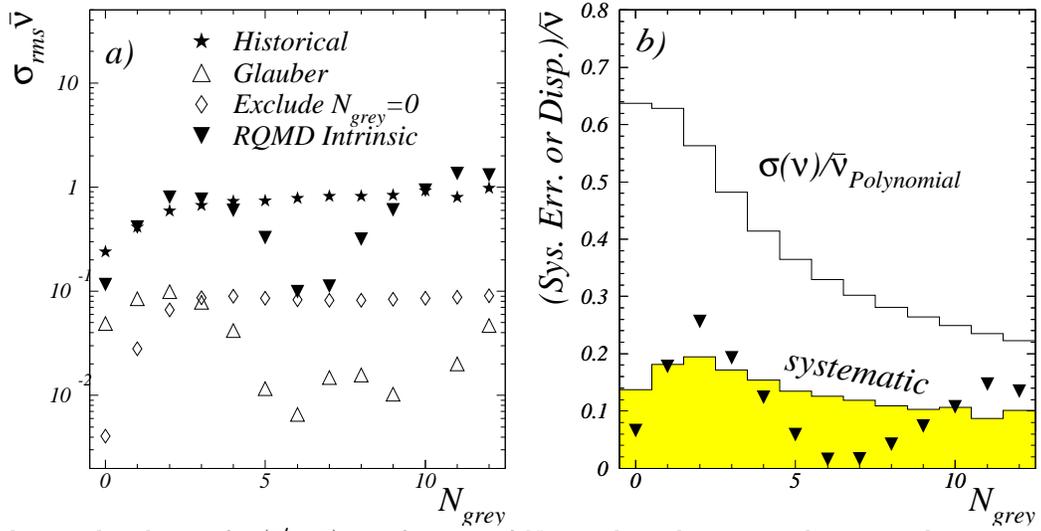}
    \caption{a) 
      the rms distribution for $(\overline{\nu}'-\overline{\nu})$ as
  a function of 
  $N_{grey}$, where the prime indicates an alternate set of systematic
  cuts, or the intrinsic value from RQMD.  b) the relative
  systematic error (shaded) for the combined one standard deviation
  errors from the three re-analyses described in the text.  For
  comparison the RQMD relative intrinsic difference (triangles) and
  the polynomial model relative dispersion (unshaded) are also shown.
      } 
    \label{fig:systematics}
  \end{center}
\end{figure}

We estimate the systematic errors through a set of re-analyses of the
polynomial model applied to p-Au data set with the following changes:
\begin{description}
\item [historical --] define $N_{grey}$ to be $0.3<p<1.0$~GeV/c for protons,
 and $0.6<p<2.0$~GeV/c for deuterons,
\item [glauber --] substitute Glauber model for Hijing in the calculation of
$\pi(\nu)$ (see Fig.~\ref{fig:nu_distributions}),
\item [exclude --] remove $N_{grey}=0$ bin from fit to data.
\end{description}
The first two modifications are straightforward alternatives to the
standard analysis.  Removing the $N_{grey}=0$ bin checks for a bias in
our interaction trigger.  Fig.~\ref{fig:systematics}a shows the rms
deviations in the extracted $\overline{\nu}(N_{grey})$ with respect
to the standard analysis.  The historical momentum cuts show the
largest discrepancy.  For comparison, the magnitude of the difference
between intrinsic $\overline{\nu}(N_{grey})^{\rm RQMD}$ and
polynomial analysis of $\overline{\nu}(N_{grey})$ for the RQMD E910
model set is also shown in this figure.  This difference,
$\delta\overline{\nu}(N_{grey})^{RQMD}$, should include all
systematic effects of this analysis in addition to systematic errors
inherent in RQMD.  The dependence of
$\delta\overline{\nu}(N_{grey})^{RQMD}$ on $N_{grey}$ should not be
taken as a true reflection of the behavior of the systematic errors,
but rather an indication of their range.  Note that it oscillates
around the rms deviation of $\overline{\nu}(N_{grey})$ for the
historical analysis which is the dominant contribution to the
systematic error.

Fig.~\ref{fig:systematics}b shows the relative systematic error for
the sum in quadrature of all three re-analyses.  The 1$\sigma$
systematic error is 10-20\%, peaking at $N_{grey}=0$ ($\nu\approx 3$)
for p-Au.  This is significantly smaller than the
dispersion, $\sigma(\nu(N_{grey}))$, shown in the figure relative to
$\overline{\nu}(N_{grey})$ for the standard analysis.  The RQMD
intrinsic difference is re-plotted as a relative difference to compare
with our final estimate of the relative systematic error.

\section{Conclusions}
\label{sec:conc} 

We have measured the slow proton and deuteron production in 18~GeV/c
proton collisions with three targets, Be, Cu, and Au in the momentum
range relevant to a determination of collision centrality.  RQMD, a
full intra-nuclear cascade model, provides reasonable agreement with
the $N_{grey}$ distribution for the p-Au data.  The simple GCM and
Polynomial models are also fit to the data as part of the procedure to
extract $\overline{\nu}(N_{grey})$.  The GCM imposes no upper bound on the
number of protons that can be emitted and therefore over-predicts the
$N_{grey}$ distributions for all targets.  Though not a
perfect fit ($\chi^2/$dof of 1--100), the Polynomial model gives a
better description of the data.

We are unable to comment directly on the applicability of the model of
Hegab and H\"{u}fner until we can compare its predictions for the
$N_{grey}$ distributions to data.  However, from the result of the
polynomial fits we conclude that there is little $\nu^2$ in the
dependence of $N_{grey}$ on centrality, contrary to the predictions
in~\cite{Heg82:PAnu,Heg81:PAnu}.  We cannot say that this contradicts
results from previous experiments.  The authors in ~\cite{Heg82:PAnu}
compare to only one non-emulsion data set (where the target is known),
and find reasonable agreement with their model.  However, a later
publication from this experiment found $\overline{N_{grey}}$ to be
approximately linear in $\overline{\nu}$~\cite{Bra82}, a result also
obtained in~\cite{Fae79}.  This evidence for a linear relation is
consistent with the results of the polynomial fit and the central
assumption of the GCM.  The exact reason for this linear dependence is
unknown, but we speculate that the main assumptions of the GCM are
approximately true: an independent and equivalent cascade for each
primary hadron-nucleon interaction.  Deviations from this could be the
reason for the presence of a finite constant term in the polynomial
analysis.

Our main result is the determination of centrality for a set of
collisions from the measured $N_{grey}$ with two different models.
The predictions of the two models differ by less than the predicted
dispersions for most $N_{grey}$.  Both models have been checked
against a full cascade model, RQMD, and the intrinsic
$\overline{\nu}(N_{grey})^{\rm RQMD}$ lies between the GCM and
polynomial results.  On the basis of the fits to the data, we ascribe
the more accurate measure of $\overline{\nu}(N_{grey})$ to the
polynomial model.  Finally, we establish a systematic error for this
centrality measure that is 10-20\% of $\overline{\nu}(N_{grey})$.

\section{Acknowledgments}

We wish to thank Dr.~R.~Hackenburg and the MPS staff, J.~Scaduto, and
Dr.~G.~Bunce of Brookhaven National Lab for their help in staging and
running the experiment.  We are particularly indebted to, Dr.~Tom
Kirk, for his support and encouragement in pursuing the scientific
program of E910.  We are also grateful to Dr.~Heinz Sorge for his
generous correspondence regarding the collision history of RQMD.
%
%Tom Kirk's official title is Associate Laboratory Director for High
%Energy and Nuclear Physics.
%

This work has been supported by the U.S. Department of Energy under
contracts with BNL (DE-AC02-76CH00016), Columbia University
(DE-FG02-86-ER40281), LLNL (W-7405-ENG-48), and the University of
Tennessee (DE-FG02-96ER40982) and the National Science Foundation
under contract with the Florida State University (PHY-9523974).

\bibliographystyle{prsty} \bibliography{hi,nslow}

\begin{thebibliography}{10}
\bibitem{greynote}
This terminology originates with the identification of emulsion tracks. The
  grey tracks were those with a grain density between 1.4 and 6.8 times that of
  minimum ionizing tracks. They peak forward and are composed of protons,
  deuterons and a few percent tritons, $^4$He, and mesons. Those with still
  higher grain density were called balck, and were identified with the proton
  and nuclear fragments distributed isotropically in the target frame.
\end{thebibliography}

\pagebreak
\onecolumn

%
% Tables for preprint style
%
\begin{table}
\caption{Event statistics and forward angle cuts for all targets.
Cuts for deuterons are in parentheses.}
\label{tab:statsncuts}
\begin{tabular}{|l||c|c|c|c|} \hline
Target & $Events$ & $\cos \theta_f$ & protons & deuterons \\
\hline
Au & 35520 & 0.98 (0.97) & 56881 & 10622 \\
\hline   
Cu & 49331 & 0.98 (0.96) & 45784 & 6224 \\
\hline
Be & 100609 & 0.94 (0.94) & 30622 & 3366 \\
\hline
\end{tabular}
\end{table}

\begin{table}
\caption{Mean values for $N_{grey}$, $\nu$ and GCM fit parameters.}
\begin{tabular}{|l||c|c|c|c|c|} \hline
Target & $\overline{N}_{grey}$ & $\overline{\nu}$ & $X = \overline{N}_{grey}/\overline{\nu}$ & $X_{\rm fit}$ & $\chi^2$/dof \\
\hline
Au & 1.98 & 3.63 & 0.353 & 0.351 $\pm$ 0.001 & $3.04\cdot 10^4/15$ \\
\hline 
Cu & 1.06 & 2.40 & 0.306 & 0.306 $\pm$ 0.001 & 910/12 \\
\hline
Be & 0.342 & 1.36 & 0.201 & 0.201 $\pm$ 0.001 & 4007/6 \\
\hline
\end{tabular}
\label{tab:andfit}
\end{table}

\begin{table}
\caption{Coefficients for polynomial fit to $N_{grey}$.}
\begin{tabular}{|l||c|c|c|c|} \hline
Target & $c_0$ & $c_1$ & $c_2$ & $\chi^2$/dof \\
\hline
Au & -0.27 $\pm$ 0.02 & 0.63 $\pm$ 0.01 & -0.0008 $\pm$ 0.0012 &  1639/13 \\
\hline  
Cu & -0.17 $\pm$ 0.02 & 0.51 $\pm$ 0.02 & -0.00005 $\pm$ 0.00242 & 15/10 \\
\hline
Be & -0.075 $\pm$ 0.008 & 0.306 $\pm$ 0.006 & --- & 95/5 \\
\hline
\end{tabular}
\label{tab:polyfit}
\end{table}

\begin{table}
\caption{Coefficients for polynomial fit to $N_{grey}$ with $c_0$
constrained to be zero.}
\begin{tabular}{|l||c|c|c|c|} \hline
Target & $c_0$ & $c_1$ & $c_2$ & $\chi^2/{\rm dof}$ \\
\hline
Au & --- & 0.439 $\pm$ 0.006 &  0.019 $\pm$ 0.001 & $1.06\cdot 10^4/15$ \\
\hline  
Cu & --- & 0.369 $\pm$ 0.005 &  0.021 $\pm$ 0.001 & 53/11 \\
\hline
Be & --- & 0.206 $\pm$ 0.004 &  0.026 $\pm$ 0.003 & 61/5 \\
\hline
\end{tabular}
\label{tab:polyfit2}
\end{table}

\begin{table}
\caption{Coefficients for polynomial and GCM fits to $N_{grey}$ from
RQMD 18~GeV/c p-Au.}
\begin{tabular}{|l||c|c|c|c|} \hline
Target & $c_0$ or X & $c_1$ & $c_2$ & $\chi^2$/dof \\
\hline
GCM E910     
   & 0.3605$\pm$0.0006 & --- & --- & 8300/16 \\
\hline
GCM 4$\pi$ 
   & 0.4933$\pm$0.0006 & --- & --- & 6909/23 \\
\hline
Polynomial E910 
   & -0.136$\pm$0.009 & 0.663$\pm$0.006 &  -0.0089$\pm$0.0007 & 80/14 \\
\hline  
Polynomial 4$\pi$ 
   & -0.443$\pm$0.007 & 1.155$\pm$0.007 & -0.0075$\pm$0.0008 & 69/21 \\
\hline
\end{tabular}
\label{tab:rqmd_ngrey_fits}
\end{table}

%
% Figures for preprint style
%

\end{document}